\documentclass[sigconf]{acmart}

\usepackage{booktabs} % For formal tables

\setcopyright{acmcopyright}
\acmArticle{4}
\acmPrice{15.00}

\usepackage{amsmath}
\usepackage{graphicx}
\usepackage{listings}
\usepackage[utf8]{inputenc}
\usepackage{booktabs} 
\usepackage{xcolor}
\usepackage{makecell}
\usepackage{caption}
\definecolor{eminence}{RGB}{255,0,0}

\lstdefinestyle{sparql}{
	morekeywords={PREFIX,@PREFIX,REGEX,STRAFTER,STR,STRSTARTS, NOT, EXISTS, SELECT,OPTIONAL,FROM,DISTINCT,a,WHERE,FILTER,GROUP,GROUP_CONCAT,MINUS,ORDER,LIMIT,BY,IN,AS,COUNT, ASK},
	morecomment=[l][\itshape\color{olivegreen}]{\#},
	morestring=[b][\color{blue}]\",
	sensitive=true,
	comment=[l][\itshape]{\#}, 
	backgroundcolor=\color{white},
	basicstyle=\ttfamily\footnotesize,
	numbers=left,
	numberstyle=\scriptsize\ttfamily,
	captionpos=b,
	frame=tb
}
\copyrightyear{2023} 
\acmYear{2023} 
\setcopyright{rightsretained} 
\acmConference[SAC '23]{The 38th ACM/SIGAPP Symposium on Applied Computing}{March 27-March 31, 2023}{Tallinn, Estonia}
\acmBooktitle{The 38th ACM/SIGAPP Symposium on Applied Computing (SAC '23), March 27-March 31, 2023, Tallinn, Estonia}\acmDOI{10.1145/3555776.3578739}
\acmISBN{978-1-4503-9517-5/23/03}

\begin{document}
\title{DISO: A Domain Ontology for Modeling Dislocations in Crystalline Materials}
% \titlenote{Produces the permission block, and
%   copyright information}
% \subtitle{Extended Abstract}
% \subtitlenote{The full version of the author's guide is available as
%   \texttt{acmart.pdf} document}
  
\renewcommand{\shorttitle}{Dislocation Ontology (DISO)}

\author{Ahmad Zainul Ihsan}
\orcid{0000-0002-1008-4530}
\affiliation{%
  \institution{Forschungszentrum Jülich, Institute of Advanced Simulation --
	Materials Data Science and Informatics (IAS-9), Germany}
}
\email{a.ihsan@fz-juelich.de}

\author{Said Fathalla}
\orcid{0000-0002-2818-5890}
\affiliation{
  \institution{Forschungszentrum Jülich, Institute of Advanced Simulation --
	Materials Data Science and Informatics (IAS-9), Germany}
  \institution{Faculty of Science, University of Alexandria, Egypt}
}
\email{s.fathalla@fz-juelich.de}

\author{Stefan Sandfeld}
\orcid{0000-0001-9560-4728}
\affiliation{
  \institution{Forschungszentrum Jülich, Institute of Advanced Simulation --
	Materials Data Science and Informatics (IAS-9), Germany}
  \institution{RWTH Aachen University, Faculty 5, Germany}	
	}
\email{s.sandfeld@fz-juelich.de}

% The default list of authors is too long for headers}
\renewcommand{\shortauthors}{A. Z. Ihsan et al.}

\begin{abstract}
%  Materials Science and Engineering (MSE) is mainly concerned with the design, synthesis, properties, and performance of materials. 
Crystalline materials, such as metals and semiconductors, nearly always contain a special defect type called dislocation.
This defect decisively determines many important material properties, e.g., strength, fracture toughness, or ductility.
Over the past years, significant effort has been put into understanding dislocation behavior across different length scales via experimental characterization techniques and simulations. 
%% While there is still a lack of a common standard to represent the dislocation structures and connect to other related domains.
% Domain ontologies play an important role in describing concepts in a particular domain of interest.
%% a common foundation to enable knowledge representation and data interoperability, which are important components to establish FAIR principles. 
This paper introduces the dislocation ontology (DISO), which defines the concepts and relationships related to linear defects in crystalline materials. 
%% Subsequently, the development of DISO follows several practices, e.g., using well-known ontology metadata and reusing existing ontologies. 
We developed DISO using a top-down approach in which we start defining the most general concepts in the dislocation domain and subsequent specialization of them. 
DISO is published through a persistent URL following W3C best practices for publishing Linked Data.
Two potential use cases for DISO are presented to illustrate its usefulness in the dislocation dynamics domain.
The evaluation of the ontology is performed in two directions, evaluating the success of the
ontology in modeling a real-world domain and the richness of the ontology.
% We demonstrated the effectiveness of the ontology by showing its ability to answer several predefined queries and.
% Lastly, the evaluation of ontology is shown by answering the competency questions by means of SPARQL query.
\end{abstract}

%
% The code below should be generated by the tool at
% http://dl.acm.org/ccs.cfm
% Please copy and paste the code instead of the example below. 
%


\maketitle

\section{Introduction}
\label{sec1/introduction}
A dislocation is a one-dimensional lattice defect responsible for plastic deformation in metals and other crystalline materials. 
It plays an essential role in determining mechanical properties such as strength and ductility. 
Since the existence of the dislocation was postulated in the 1930s by Orowan~\cite{orowan1934}, Taylor~\cite{taylor1934a}, and Polanyi~\cite{polanyi1934}, significant efforts have been undertaken to understand all details of systems of dislocations based on, e.g., dedicated microscopy techniques for visualizing dislocations. 
These techniques and various simulation methods were developed to predict the evolution of dislocation.

In the recent years, data-driven approaches have brought new methods and tools for analyzing and understanding the evolution of dislocation systems~\cite{sandfeld2015,salmenjoki2018,steinberger2020}. 
Similarly, the whole MSE field, a parent domain of the specialized domain of dislocations, is undergoing an unprecedented change. 
This change brings simulations and experiments together and ultimately makes digital transformation in the field of Materials Science and Engineering (MSE) possible~\cite{Prakash2018_PrM55,EMMC-MatDig,Kimmig2021}.

While dislocations can be directly represented through the positions of atoms, in many cases, the idealized representation of the dislocation as a mathematical line is preferred as it allows to consider larger systems with fewer degrees of freedom. The ``scale'' on which these mathematical lines are defined is larger than the atomic scale (atoms are not visible there) but still contains effective atomic scale information in the form of the dislocation microstructure. Therefore, this is often called the ``mesoscale'' even though this is not a well-defined notion.
To fully understand the behaviour of materials, aspects from different length scales need to be considered.
This makes the knowledge representation of systems of dislocations challenging, even though so far this has not been perceived as a significant research hindrance in materials science. 
Formal knowledge representation, i.e., via ontologies, enables interoperability and data handling between related MSE domains, thus enabling machine actionability.
Ontology also allows the domain knowledge to be represented by logical axioms that the machine can understand.
% Furthermore, ontology is explicit, i.e., the meaning of all concepts is defined, and ontology is shared by a common consensus. 
Ultimately, an ontology is an important component to enable the establishment of the FAIR principles (Findable, Accessible, Interoperable, and Reusable)~\cite{wilkinson2016} in the MSE domains.

% The initial work representing the knowledge of dislocation was started in the previous publication~\cite{ihsan2021_1}. 
% While in the previous publication, the domain descriptions and abstract conceptualization were mainly explained. 
% The implementations in ontology development are still lacking. 
This paper represents the dislocation ontology (DISO), an ontology representing the concepts and relationships in linear defects in crystalline materials. 
DISO is an extended version of the ontology presented in Ihsan et al.~\cite{ihsan2021_1}.
Indeed, DISO is a part of the Dislocation Ontology Suite (DISOS)\footnote{\url{https://purls.helmholtz-metadaten.de/disos}}. 
DISOS is an ontology suite comprising several modules describing materials scientific concepts, representations of dislocations, and different simulation models in the dislocation domain.
The contributions can be summarized as follows:

\begin{itemize}
    \item covering further characteristics of dislocation, including the mathematical and numerical representation of the dislocation line, 
    \item adding new concepts, including Bravais Lattice, Space Group, Point Group, Crystal System, and Vector,
    % , representation in the Crystal Structure Ontology (CSO).
    % \item Add ontology formalization and evaluation.
    \item demonstrating the usefulness of the new version by showing its ability to answer a set of predefined queries in various complexity,
    \item creating human-readable documentation for the ontology as well as adding ontology metadata (e.g. contributor, title, date created, etc.), which will greatly improve its findability and reusability,
    \item presenting two potential use cases for DISO: Dislocation dynamics data and dislocation experiment data, 
    \item creating RDF dataset of dislocation structure data produced by MoDELib software~\cite{Po2014_1} by mapping data stored in Hierarchical Data Format 5 (hdf5)\footnote{\url{https://www.hdfgroup.org/solutions/hdf5/}} files to RDF triples using DISO classes, and
    \item evaluating the ontology using two  different evaluation strategies.
    
\end{itemize}
 
% This work introduces the dislocation ontology (DISO), which represents the concepts related to linear defects in crystalline materials and the relationships between them. Moreover, the idealization of dislocation as mathematical and numerical lines and other details of crystalline materials are also described. 

DISO is developed and maintained on a GitHub repository\footnote{\url{https://github.com/Materials-Data-Science-and-Informatics/Dislocation-Ontology-Suite/tree/main/DISO}}.
The ontology is available (in several RDF serializations) via a persistent identifier (i.e., {\url{https://purls.helmholtz-metadaten.de/disos/diso}}) provided by PIDA (Persistent Identifiers for Digital Assets)\footnote{\url{https://purls.helmholtz-metadaten.de/}}.
PIDA is a service that provides persistent identifiers for digital assets.
It employs content negotiation~\cite{berrueta2008cooking}, which is a mechanisms defined as a part of HTTP serve different versions of a resource (i.e. the HTML documentation or an RDF representation) at the same URI in accordance with the client's request. 
The dereferenceability of the IRIs of the ontology terms over the HTTP protocol (cf.\ \cite{lewis2007dereferencing}) has been checked using a Linked Data validator (i.e. Vapour\footnote{{\url{http://linkeddata.uriburner.com:8000/vapour?}}}).
Furthermore, DISO is syntactically validated by the W3C RDF validation service\footnote{\url{https://www.w3.org/RDF/Validator/}} to conform with the W3C RDF standards. 
To make it easier to understand and reuse our ontology, human-readable documentation of the ontology is generated and can be found online via its URI.
% \footnote{\url{https://materials-data-science-and-informatics.github.io/dislocation-ontology/}}.

The rest of this paper is organized as follows:
Section \ref{sec2/related_work} presents related work of materials science domain ontologies and points out the existing gaps. 
Section \ref{sec3/linear-defects} describes the physical aspects of dislocations and in section \ref{sec4/ontology-development}, the development of the ontology is described. 
Section \ref{sec5/potential-use-cases} shows two potential use cases and section \ref{sec6/evaluation} presents the evaluation of DISO using a criteria-based strategy. 
Finally, section \ref{sec6/conclusion} concludes and sketches the envisioned future work.

% PID dislocation ontology= https://purls.helmholtz-metadaten.de/diso\#className  
% PID crystal structure ontology= https://purls.helmholtz-metadaten.de/cso

% new added contributions:
% \begin{itemize}
%     \item adding the mathematical and numerical representation of the dislocation line 
%     \item add classes regarding vector, vector components, and basis for Crystal Structure Ontology
%     \item add classes regarding the space group, point group, and crystal system for Crystal Structure Ontology
%     \item add more instances, e.g., dislocation microstructure from Modelib, crystal structure, slip plane, and slip direction to the dataset (ttl file)
%     \item ontology metadata
%     \item HTML documentation
%     \item evaluation
%     \item description logics and Semantic Web Rule Language (SWRL) rules, which enable automatic reasoning on crystalline materials data
% \end{itemize}

% \newline
% \vspace{1em}

\section{Related work}
\label{sec2/related_work}
In the past decades, several researchers have been involved in the knowledge representation of various fields of science~\cite{fathalla2020towards}, including Materials Science and Engineering (MSE), through developing domain ontologies.

Plinius~\cite{plinius1994} is an ontology developed for describing ceramic materials covering the conceptualization of chemical compositions ranging from the single atom to complex chemical substances. 
In the ``Materials Ontology''~\cite{ashino2010}, a detailed structure of materials information consisting of substances, processes, environment, and properties was conceptualized. 

The ``\textit{International Union of Crystallography}''\footnote{\url{https://www.iucr.org/}} published and distributed the Crystallographic Information File (CIF) and the further developed CIF2 which currently co-exists with the CIF file format. 
CIF serves as a \emph{de facto} standard for crystallographic information format exchange. 
The authors of CIF in \cite{hall2016} also have further developed the STAR/CIF Ontology that was written in the mathematical symbolic script language called dREL~\cite{drel2012}. 

Another ongoing effort to establish semantic standards that apply at the highest possible level of abstraction is the \textit{Elementary Multi-perspective Material Ontology} (EMMO)\footnote{\url{https://emmo-repo.github.io}}. 
EMMO has been developed in the context of the \textit{European
Materials Modelling Council}\footnote{\url{https://emmc.eu/}} (EMMC) and provides a common semantic framework for describing material models, characterization, and data with the possibility of extension and adaptation to other domains. 
% For instance, the application of EMMO in the mechanical testing domain was demonstrated in \cite{morgado2020}, and challenges of ontology alignment of level-domain ontologies and EMMO were addressed in \cite{horsch2020}. 
However, EMMO currently contains only a small number of sub-domains.
Furthermore, it does not include the domain of dislocations which is of great importance for materials scientists in the context of mechanical deformation of nano- and micometer sized specimens.
    
Recent development work in materials design (e.g., with regards
to ab-initio methods) and crystallographic information resulted in the \textit{Materials Design Ontology} (MDO)~\cite{li2020} and the \textit{Crystal Structure Ontology} (CSO)\footnote{\url{https://purls.helmholtz-metadaten.de/disos/cso}}, respectively.
The former represents information on the atomic structure of materials via its “structure” ontology module.
The latter represents crystallographic information needed to describe the crystallographic defect. 
However, both do not explicitly represent the physical and conceptual aspects of the crystal structure that are directly related to the representation of defects in
crystals, particularly dislocations.
    
In summary, it can be concluded that even though significant progress has been made concerning the ontology design in a number of related domains, it becomes clear that a level-domain ontology of dislocations in crystalline materials is still missing. 
% Furthermore, there is still a gap between 
% \Stefan{The gap should be between TWO things/aspects -- which are they? This last paragraph needs to be rephrased to make it strong.}
% the existing crystal structure ontologies regarding the representation of crystalline defects. 
Furthermore, there are several attempts to represent materials science phenomena in the literature, even though few researchers have addressed the problem of semantically representing crystalline defects. 
Therefore, developing a domain ontology of dislocations in crystalline materials fills a clear gap and is an important step for the domain of materials science.

    \section{Description of the Domain}
    \label{sec3/linear-defects}
\begin{figure}[bt]
    	\includegraphics[width=\linewidth, height=2.5cm]{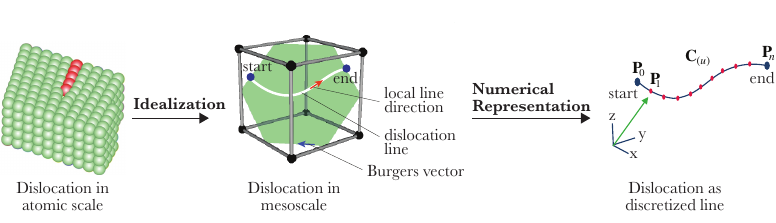}
    	\caption[]{The idealization represents the dislocation in the mesoscale. Here, the individual atoms are no longer visible. This idealization reduced the tube-like defect ``region'' to a mathematical line. On the rightmost, the numerical representation of a mathematical line as the discretized line is illustrated.}
    	\label{fig/sec3:idealization-dislocation}
    \end{figure}

    In real crystalline materials, atoms are typically not perfectly ordered or positioned -- at least not everywhere. 
    Typically, various types of crystallographic defects in a piece of material destroy the local order of the crystal structure (on top of thermal fluctuations that also affect the atomic positions).
    Such a defect might be a point defect (e.g., a missing or extra atom) or, as shown in the left panel of \autoref{fig/sec3:idealization-dislocation}, a dislocation (a strongly localized, tube-like region of disorder containing the highly disordered dislocation core at the center). 
    
    According to the geometry, there are two types of dislocations, the so-called ``edge'' and ``screw'' dislocation. The latter is characterized by having a line sense parallel to its Burgers vector, whereas in the former, the line sense is perpendicular to its Burgers vector.
    Furthermore, also the atoms in the vicinity of the dislocation are displaced from their perfect lattice sites. 
    This lattice distortion subsequently results in a stress field in the crystalline material around the dislocation.

    % While in general, the notion of “defect” has asomewhat negative connotation. This deviation from the perfect structure/orderresults in important properties, e.g., electrical, mechanical, or thermal. In thefollowing,  the  one-dimensional  defect  type  is  considered;  other  defects  will  bediscussed in a follow-up publication.

    % In the context of plastic deformation, a \emph{dislocation} is defined as the boundary of a slipped area within which atoms are displaced by the size of an elementary unit translation given by the so-called \emph{Burgers vector}. However, the question arises on which granularity level a dislocation should be defined? One can define it in terms of displaced atoms, or -- as will be done in the following (see \autoref{fig/sec3:idealization-dislocation}) -- 
    
    From a \emph{mesoscopic} point of view, details from the smallest considered scale, the scale of the individual
    atoms are no longer visible.
    %One can take a \emph{mesoscopic} view as an idealization. 
    %In this view, the individual atoms are no longer visible.
    Nonetheless, the dislocation line still can be observed: the tube-like defect ``region'' is reduced to an idealized mathematical line as shown in \autoref{fig/sec3:idealization-dislocation} (see the center figure). 
    Thus, moving from the atomic scale to such a mesoscale is an idealization and strongly reduces the amount of information to only the utmost necessary information.

    A wide range of experimental and computational techniques are used to observe and predict dislocations in crystalline materials. E.g., on the atomic scale, field ion microscopy or high-resolution transmission electron microscopy is used to image the arrangement of atoms.
    Then on the mesoscale, these techniques determine the properties of individual dislocations and study the arrangement, distribution, and density of dislocations in crystalline materials. 
    Examples of the techniques at the mesoscale are Transmission Electron Microscopy (TEM) and, as a simulation method, Discrete Dislocation Dynamics (DDD).

    TEM is a microscopy technique that uses a particle beam of electrons, transmitted through a specimen and passing some lenses, ultimately generating a higher resolution (highly-magnified) image. 
    The transmitted beam is deflected once it hits the dislocation. 
    The deflection reduces the intensity of the transmitted beam and increases the intensity of the diffracted beam. 
    The dislocation then appears as a dark line in the bright-field image. 
    % Note that the crystalline material specimen is subjected to a thin section. This thin section facilitates the electrons to pass through the specimen.

    The motion and interaction of dislocation lines during plastic deformation mainly follow the governing equations of (linear) elasticity theory. The DDD simulation method treats dislocations as mathematical lines represented as polygons or splines and moves them according to elastic interactions and further ``local rules''. 
    % DDD allows to study of the mechanical properties of the crystalline material as these are ultimately the result of dislocation motion and interaction, governing the plastic deformation. 
    
    %   \todo{\Stefan{Same procedure again: at least the tiny labels in the middle need a larger fontsize...but this figure is very good now, Ahmad!}
    %   \Ahmad{Done..}}
    
    Both techniques mentioned above observe the dislocation as a mathematical line. 
    However, they apply different approaches to represent the mathematical line. 
    In TEM, the image containing the dislocation lines is represented by a pixels image. 
    % Pixels containing a part of the dislocation lines have a lower RGB (darker color) value in a bright-field image.
    In DDD, the mathematical dislocation line needs to be discretized to solve the governing equations numerically. 
    The particular type of discretization represents the mathematical line through the \emph{numerical representation}. 
    The discretization steps~\cite{ghoniem1999} are (also see the rightmost figure in  \autoref{fig/sec3:idealization-dislocation}): the oriented curve of the dislocation line, $\mathbf{C}(u)$, where $0\leq u \leq 1$, is discretized into a number of segments, $n_s$.
    The position vector for any point on the segment is defined by $\mathbf{C}(u) = \mathbf{P}(u)$ and the shape of a segment is defined by choosing the \emph{shape function}, $\mathbf{N}_i(u)$, e.g., linear, circular, cubic spline segment. However, this is only one possible discretization type, and also others are commonly used in the DDD simulation community.
    % Suppose a set of generalized coordinates, $\mathbf{q}_i^{(j)}$, where \emph{(j)} is a segment, and linear shape function are chosen, then the position vector equation of the segment is written as 
    % \begin{equation}
    %     \mathbf{C}^{(j)}(u) = \mathbf{q}_i^{(j)}\mathbf{N}_i(u)
    %     \label{eq/discretized}
    % \end{equation}
    % where the shape funtions are given by
    % \begin{equation}
    %     N_1(u) = 1-u,\ \ N_2(u) = u
    %     \label{eq/linear_shape_func}
    % \end{equation}

\section{Ontology Development}
\label{sec4/ontology-development}

Developing domain ontologies attracts considerable interest from several private and public sector organizations to capture their knowledge about their domain of interest in a form that machines can understand. 
In fact, the main goal of developing ontologies is to share conceptualizations through which humans express meanings of things that can act as communication interfaces between humans and machines. 
% This greatly supports research data publishing in a form that machines can easily read, whether computers, IoT devices, or mobile phones. 

\begin{figure*}[htb]
% \vspace{-3.5mm}
\centering
\includegraphics[width=0.8\textwidth]{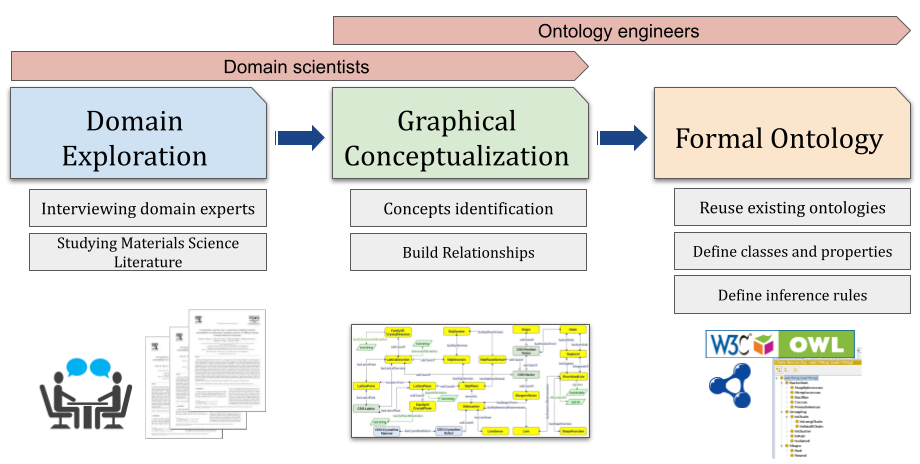}
\caption{The workflow of the dislocation ontology development, illustrating the main phases, subprocesses, and roles involved in the whole process.}
\label{fig:workflow}
% \vspace{-3.0mm}
\end{figure*}

This section describes the development process of DISO, which is an iterative approach, which means we started with an initial version and then revise and refine the evolving ontology. 
\autoref{fig:workflow} illustrates the workflow of this process which is inspired from  several well-known ontology development methodologies~\cite{kotis2020ontology}. 
This figure presents the main phases, their sub-tasks, and the roles involved in the whole process.
We have frequently interviewed domain scientists as well as ontology engineers during the whole process in order to improve the final outcome.
This continues through the entire development process.
% In the formal ontology step, OWL2 DL is used as the formal representation language of DISO.

% As shown in the figure, the Web Ontology Language (OWL) is used to implement the ontology.

\subsection{Ontology Metadata}
\label{sec:overview}

\begin{sloppypar}
An important step towards improving ontology reuse is to provide a systematic and comprehensive description of ontologies~\cite{Simperl2011-Ontology-metadata}, i.e., ontology metadata. 
Often, ontologies are not well documented or miss information about the ontology itself. 
Consequently, these ontologies are therefore suffering from several problems, including 1) being not optimally accessible to potential users, 2) hindering their reusability, and 3) identifying required ontologies for specific applications efficiently. 
Accordingly, several DCMI Metadata Terms\footnote{{\url{https://www.dublincore.org/specifications/dublin-core/dcmi-terms/}}} are added to the ontology, involving \texttt{terms:contributor},  \texttt{terms:created},  \texttt{terms:title}, and \texttt{vann:preferredNamespacePrefix}. 
As a next step, providing human-readable documentation for the ontology is also crucial because it is one of the main reuse problems identified in \cite{fernandez2019ontologies}.
\end{sloppypar}

\subsection{Reuse of existing models}
\label{sec:reuse}
One of the initial steps in the ontology development is to reuse terms (i.e., classes or properties) from existing ontologies that describe the same domain or topic. 
However, it is challenging for ontology engineers to decide which of the existing ontologies is suitable to be reused. 
In fact, the more ontologies a model reuses, the higher the value of its semantic data is. 
Therefore, reusing terms from other ontologies greatly
% \Stefan{to what does 'it' refer? Please rephrase.} 
increases semantic data value~\cite{seo2019} of the developed ontology. 

% After studying the literature, we discovered that several related concepts were missing, such as \textit{crystal structure}, \textit{Bravais lattices}, \textit{point defect}, and \textit{grain boundary}. 
% Therefore, we took the step to define these concepts by developing two related ontologies, namely, the \textit{Crystal Structure ontology} (CSO) and the \textit{Crystalline Defect ontology} (CDO). 
After studying the literature, we decided to reuse several concepts from two related ontologies, namely,  \textit{Crystal Structure ontology} (CSO) and \textit{Crystalline Defect ontology} (CDO). 
The former is an ontology developed to represent crystallographic information needed to describe the dislocation. 
The latter is an ontology connecting the physical materials entity to the crystal structure and several defects in crystalline materials, e.g., point defect, dislocation, and grain boundary. 

\begin{sloppypar}
\emph{Crystalline Defect Ontology (CDO)}\footnote{{\url{https://purls.helmholtz-metadaten.de/disos/cdo}}}. In CDO, the \texttt{EMMO:Material}, which is a class from EMMO\footnote{\url{https://emmo-repo.github.io}}, is reused to describe the physical entity of crystalline materials. The \texttt{CDO:CrystallineMaterial} class is subsequently a subclass of \texttt{EMMO:Material}. 
% \autoref{fig:cdoClasses} presents the core concepts in CDO.
\end{sloppypar}

% \begin{figure}[bt]
% % \vspace{-3.5mm}
% \centering
% \includegraphics[width=\linewidth]{fig/CDO-classes.png}
% \caption{Core concepts in the CDO ontology. }
% \label{fig:cdoClasses}
% \end{figure}

\begin{sloppypar}
\emph{Crystal Structure Ontology (CSO)}\footnote{{\url{https://purls.helmholtz-metadaten.de/disos/cso}}}. In CSO, several MDO~\cite{li2020} classes are reused to describe the crystal coordinate system, motif (an arrangement of chemical species in the crystal structure) in a crystal structure, point groups, and space groups. 
The standard coordinate system is defined by \texttt{MDO:Basis} and \texttt{MDO:CoordinateVector} classes that CSO reuses. 
The motif reuses \texttt{MDO:Occupancy}, \texttt{MDO:Site}, and \texttt{MDO:Species}. 
Subsequently, to define the point groups and space groups of a crystal structure, \texttt{MDO:PointGroup} and \texttt{MDO:SpaceGroup} are reused. 
Lastly, to define the unit quantity of a property in CSO, CSO reuses several classes from QUDT (Quantities, Units, Dimensions and Data Types Ontologies) ~\cite{hodgson2014qudt}. Classes that are reused from QUDT are the \texttt{QUDT:Quantity}, \texttt{QUDT:QuantityKind}, and \texttt{QUDT:QuantityValue}. 
% \autoref{fig:csoClasses} presents the core concepts in the CSO.
\end{sloppypar}

% \begin{figure}[tb]
% % \vspace{-3.5mm}
% \centering
% \includegraphics[width=\linewidth]{fig/CSO-classes.png}
% \caption{Core concepts in the CSO ontology. }
% \label{fig:csoClasses}
% \end{figure}
% \Stefan{Can you try to move everything a bit around and make the image smaller in horizontal direction. The font is barely readable right now.}
 \begin{figure*}[tb]
 \centering
     \includegraphics[width=\linewidth]{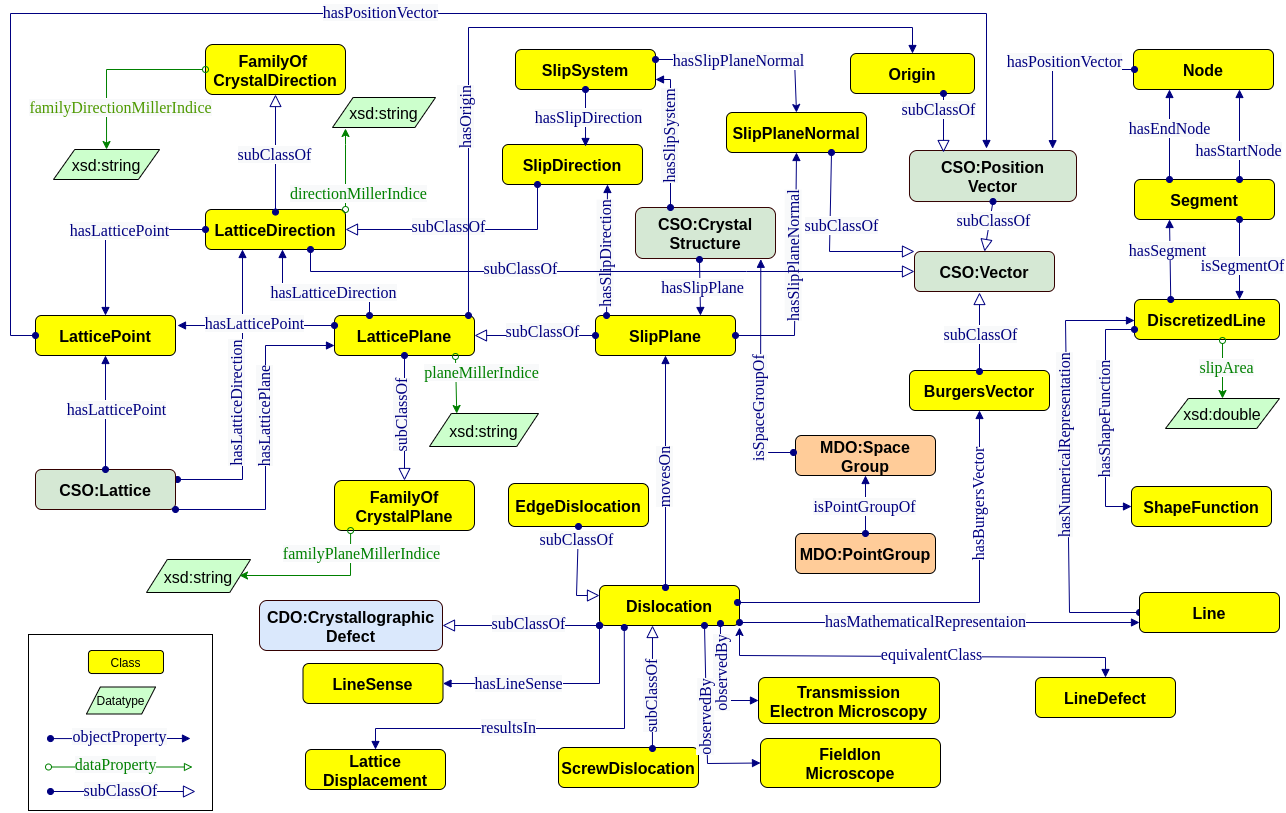}
     % trim={<left> <lower> <right> <upper>}
     %\includegraphics[trim={-2cm 0 -2cm 0},clip]{fig/seo_graph3.pdf}
     \caption{Core concepts and interconnected relationships in the DISO ontology. Arrows with open arrow heads denote \texttt{rdfs:subClassOf} properties between classes. Regular arrows visualize \textit{rdfs:domain} and \textit{rdfs:range} restrictions on properties. Furthermore, colored boxes represent different ontologies, e.g., MDO, CDO, CSO, and DISO.}
     \label{fig:mainConcepts}
 \end{figure*}

\subsection{Main Classes}
\label{sec4/main-classes}

The classes in our ontology are divided into two sets; 1) imported from the ontologies introduced in \autoref{sec:reuse} and 2) newly defined classes that are not explicitly defined in the existing related ontologies. 
\begin{sloppypar}
\emph{Imported classes}.
Several classes have been imported from the crystal structure ontology (CSO), involving  \texttt{CSO:Lattice} which represents the periodic arrangement of one or more atoms, \texttt{CSO:Vector} which represents quantities that have both magnitude and direction, and imported from crystallographic defect ontology (CDO), involving \texttt{CDO:CrystallographicDefect} which represents lattice irregularity/lattice defects by having one or more of its dimensions on the order of an atomic diameter.
\end{sloppypar}

\emph{New classes}.
Here, the focus is put on some chosen classes of the main classes in the crystalline materials and line defect: 
% \begin{itemize}[topsep=0pt,parsep=0pt,partopsep=0pt,leftmargin=10pt,labelwidth=6pt,labelsep=4pt]
  1) \texttt{Dislocation}, as the entity of main interest, which models a linear or one-dimensional defect around which some of the atoms are displaced,
  2) \texttt{SlipPlane}, which models the lattice plane to which the dislocation is constrained to move,
  3) \texttt{SlipDirection}, which models the lattice direction where the slip occurs in the crystalline materials,
  4) \texttt{LatticePlane}, which models the lattice plane where it forms an infinitely stretched plane that cuts through lattice points.
  5) \texttt{LatticeDirection}, which models the direction inside the lattice that connects two lattice points, and
  6) \texttt{DiscretizedLine}, which models the numerical representation of the dislocation line as a mathematical line, e.g. an oriented curve, that is  discretized into a number of segments.
  
\subsection{Properties}
\begin{sloppypar}
% In the same way as explained in \autoref{sec4/main-classes}, 
Both data and object properties in DISO are divided into two categories: newly defined properties and reused ones:

\emph{New properties}.
The relationships between concepts are implemented as object properties, for example, the relationship between \texttt{Dislocation} and  \texttt{TransmissionElectronMicroscopy} can be represented by \texttt{observedby} property.
Similarly, the relationship between \texttt{Dislocation} and \texttt{LineSense} is represented by \texttt{hasLineSense} property.
Furthermore, several data properties also have been defined, such as  \texttt{directionMillerIndice}, \texttt{planeMillerIndice}, and \texttt{slipArea}.

\emph{Reused properties}.
Property \texttt{cso:hasPositionVector} from the CSO ontology  and several data properties from DCterms for adding ontology metadata (see \autoref{sec:overview}) are reused. After defining new properties and identifying reused ones, the domain and range for each property using \texttt{rdfs:domain} and \texttt{rdfs:range} are defined, respectively. 
For instance, the domain of the data property \texttt{diso:slipArea} is \texttt{diso:DiscretizedLine} and the range is \texttt{xsd:double}.

\emph{Restricting properties}. 
% Several properties are also imposed with ``guarded'' restrictions, i.e., use universal restrictions ($\forall$) instead of ``domain'' and ``range''. 
% The guarded restrictions are commonly used by Ontology Design Patterns (ODPs) and they apply weaker ontological commitments and foster wider reuse~\cite{vardeman2017}. 
% It is because the restrictions only apply to local range constraints, e.g., $LatticePlane \sqsubseteq  \forall hasLatticeDirection.LatticeDirection$ imposes a local range of \texttt{LatticeDirection} on the property \texttt{hasLatticeDirection} for the class \texttt{LatticePlane}.
Several DISO classes use property restrictions, e.g., value constraints. 
For instances, the \texttt{observedby} property which connects \texttt{Dislocation} and \texttt{TransmissionElectronMicroscopy} is restricted by a value constraint of \texttt{owl:someValuesFrom} and  \texttt{hasBurgersVector} which connects \texttt{Dislocation} and \texttt{BurgersVector} is restricted by a value constraint of \texttt{owl:allValuesFrom}.
\end{sloppypar}

\begin{figure*}[tb]
 \centering
    	\includegraphics[width=\textwidth]{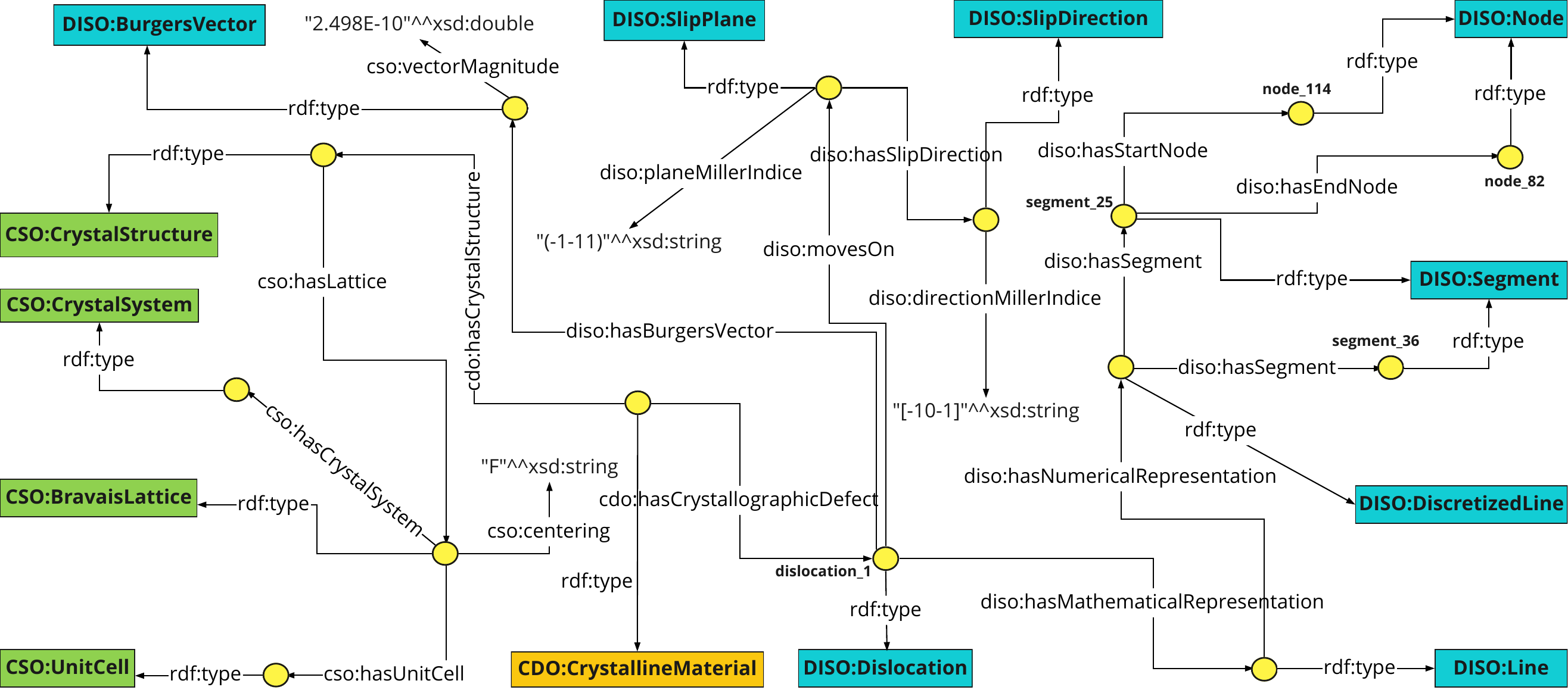}
    	\caption[]{Sample of the instances of the dislocation dynamic data. Yellow points denote individuals of classes. Each individual is defined by an arrow having \textit{rdf:type} relationship to the respective class and individuals are connected by object properties defined in DISO. Colored boxes represent classes from different ontologies, e.g., CDO, CSO, and DISO.}
    	\label{fig/instances}
\end{figure*}

\subsection{Instantiation of the Dislocation Ontology}
As shown in \autoref{fig/instances}, an excerpt of the simulation data of dislocation structure annotated by DISO is illustrated. 
% To not overflow the figure, it only contains the crystallographic information and a small portion of dislocation in the crystalline material.
% To not overflow the figure, 
Due to the space limit, we only added a sample of crystallographic and dislocation information.
We instantiate several crystallographic-related classes: crystal structure, Bravais lattice, crystal system, and unit cell. 
Furthermore, we instantiate several dislocation-related classes: dislocation, slip plane, slip direction, Burgers vector, and the representation of dislocation in numerical simulation, e.g., line, segment, and node.
Lastly, the instance of crystalline material connects the crystallographic information, i.e., crystal structure and dislocation instances.

% \vspace{0.5cm}
%   \todo{\Stefan{Same here: please try to much things around so that everything becomes better readable without having to strongly zoom in.}}
\section{Potential Use Cases}
\label{sec5/potential-use-cases}

In this section, we present two potential use cases for DISO: dislocation dynamics simulation data and dislocation experiment data. 

\textbf{Use case 1}. Dislocation dynamic (DD) simulation is a computational technique/simulation to determine individual dislocation properties and study the arrangement, distribution, and density of dislocations in crystalline materials. In the dislocation domain, there are various DD software available, including MoDELib~\cite{Po2014_1}, ParaDIS~\cite{arsenlis2007enabling}, and microMegas~\cite{devincre2011}. Each software has a different set of metadata to structure the input and output of the simulation. In this regard, annotating/mapping the DD data/metadata to DISO enables data interoperability between various DD software. In addition, it enables semantic search over materials science databases, particularly in the dislocation domain. 

\textbf{Use case 2}. An example of dislocation experiment data comes from Transmission Electron Microscopy (TEM). TEM is a microscopy technique that uses a particle beam of electrons transmitting through a specimen. In TEM, the generated data is a higher resolution (highly-magnified) image/video that observes dislocation in a crystalline material. While the TEM image/video already has some metadata to describe the data, the descriptive metadata about the observed dislocation is still lacking. In this regard, leveraging DISO along with the research data packager, such as RO-Crate~\cite{soiland2021packaging}, give the formal metadata description to the dislocation experiment data at ease, and ultimately it enables the data interoperability. Moreover, since RO-Crate is based on JSON-LD, annotating data with DISO is straightforward.

\section{Evaluation}
\label{sec6/evaluation}
% In this section, the evaluation of DISO is described. 
The evaluation of the quality of an ontology can be performed by calculating metrics that evaluate the richness of the given ontology (i.e. criteria-based evaluation)~\cite{seo2019}.
In this section, we are evaluating our ontology in two directions, evaluating the success of the ontology in modeling a real-world domain and the quality of the ontology. 
In one direction, we carried out the following steps in order to evaluate the effectiveness of DISO: 1) defining a set of competency questions (CQs) based on domain expert feedback during several interviews, 2) formulating SPARQL queries corresponding to CQs, 3) running the resultant SPARQL queries against the ontology instances (cf. \autoref{sec:data}) using a SPARQL endpoint, 4) analyzing the query results by comparing them to the correct answers given by domain experts.
On the other hand, we assess the ontology using OntoQA~\cite{tartir200} evaluation model.

\subsection{Data}
\label{sec:data}
The dislocation dynamics data produced is a virtual specimen of Nickel that consists of a number of dislocation lines. 
To fully describe the virtual specimen, the details of the crystal structure of Nickel, such as unit cell lattice parameters, point group, and space group data are taken from the Materials Project~\cite{jain2013} repository.

In order to carry out the aforementioned steps, we prepared a dislocation dynamics dataset\footnote{\url{https://github.com/Materials-Data-Science-and-Informatics/Dislocation-Ontology-Suite/blob/main/DISO/data/modelib-microstructure/modelib-nickel-microstructure.h5}} that can be used for illustrating how DISO is useful by answering competency questions prepared by domain experts.
Competency questions (CQs) determine what knowledge has to be entailed in the ontology as well as a way of evaluating how useful the ontology is by showing its capability to answer.
Therefore, we write Python scripts\footnote{\url{https://github.com/Materials-Data-Science-and-Informatics/Dislocation-Ontology-Suite/tree/main/DISO/python-script/modelib}} (using rdflib 6.0~\cite{rdflib} Python library) that map DDD MoDELib~\cite{Po2014_1} software output with DISO classes and generate RDF triples ($\sim 4K$ triples), which can be found on DISO GitHub repository.

\subsection{Competency Questions}
Evaluating ontologies with competency questions is among the most widely used types of evaluating ontologies~\cite{fathallaVA018,fathalla2019eventskg,say2020semantic,say2020ontology}.

% The domain exploration step (cf. \autoref{fig:workflow}) results in a list of competency questions (CQs). 
A total of 20 CQs (defined in the domain exploration step (cf. \autoref{fig:workflow})) were collected and categorized into three categories: the crystal structure, the dislocation, and the vector information.

For illustration, we pick CQ1 from the CQs set (listed in \autoref{tab:CQs}). The corresponding SPARQL query to CQ1 is listed in \autoref{lst:SPARQL1}. 
% instances of DISO classes (i.e., the \textit{ontology population}) from the dislocation data generated from the DDD MoDELib~\cite{Po2014_1} software. 

% The generated instances are stored in a turtle file (RDF dataset).
% by mapping the dislocation data into DISO. 
% The RDF mapping is performed via the rdflib 6.0~\cite{rdflib} python library. 
% The mapping results are subsequently evaluated by answering several competency questions (CQs) via SPARQL queries. 
% The python scripts and RDF dataset are available in the DISO GitHub repository.
\begin{table}[H]
    \centering
    \caption{Samples of competency questions.}
    \begin{tabular}{@{}cl@{}} 
 \hline
No. & Competency Question  \\
 \hline\hline
 CQ1 & What are the slip systems of a given crystal structure?\\
 CQ2 & What are the slip planes of a given crystal structure? \\
 CQ3 & In which slip planes does a dislocation move on?\\
 CQ4 & What is the Burgers vector of the dislocation?\\
 CQ5 & \makecell[l]{What is the Burgers vector magnitude of the dislocation?}\\
 CQ6 & \makecell[l]{Given a slip plane of a crystal structure, what is the slip \\direction?}\\
 CQ7 & \makecell[l]{What is the family of slip direction, given a slip direction \\in the crystal?}\\
  .. & ...\\
 CQ20 & Given a Burgers vector, what is its unit?\\
 \hline
\end{tabular}
\label{tab:CQs}
\end{table} 

\begin{lstlisting}[ showstringspaces=false,language=html,basicstyle=\scriptsize\ttfamily,keepspaces=true, caption={SPARQL query corresponding to CQ1.},label={lst:SPARQL1}, frame=single]
PREFIX diso: <https://purls.helmholtz-metadaten.de/disos/diso#>
PREFIX cso: <https://purls.helmholtz-metadaten.de/disos/cso#> 
PREFIX ex: <http://example.org/>
SELECT ?cryst_struc ?slip_sys ?slip_plane_normal_val 
?slip_direction_val WHERE{
?crystal_structure a cso:CrystalStructure ; 
    diso:hasSlipSystem ?slip_system . 
?slip_system diso:hasSlipPlaneNormal ?slip_plane_normal ; 
	diso:hasSlipDirection ?slip_direction.
?slip_plane_normal diso:directionMillerIndice ?slip_plane_normal_val. 
?slip_direction diso:directionMillerIndice ?slip_direction_val.
}
\end{lstlisting}

\autoref{tab:CQ1} shows the result of executing the above SPARQL query on the generated RDF data in which the same crystal structure consists of several slip systems.

\begin{table}[h]
    \centering
    \caption{The result of CQ1 in \autoref{lst:SPARQL1}}
    \begin{tabular}{@{}cccc@{}} 
 \hline
 Crystal Structure & Slip System & Plane Normal & Slip Direction  \\ [0.5ex] 
 \hline\hline
 cryst\_struc\_0 &	slip\_sys\_0 &	[-1-11] & [0-1-1] \\
 \hline
  cryst\_struc\_0 &	slip\_sys\_1 &	[-11-1] & [110] \\
 \hline
  cryst\_struc\_0 &	slip\_sys\_2 &	[-111] & [-10-1] \\
 \hline
  cryst\_struc\_0 &	slip\_sys\_3 &	[1-1-1] & [01-1] \\
 \hline
 cryst\_struc\_0 &	slip\_sys\_4 &	[111] & [1-10] \\
 \hline
 cryst\_struc\_0 &	slip\_sys\_5 &	[-1-11] & [-10-1] \\
 \hline
 cryst\_struc\_0 &	slip\_sys\_6 &	[11-1] & [-101] \\
 \hline
cryst\_struc\_0 &	slip\_sys\_7 &	[-11-1] &	[011] \\
 \hline
\end{tabular}
\label{tab:CQ1}
\end{table}

% According to an expert feedback, a slip system consists of two vectors of the slip plane normal and the slip direction that are orthogonal to each other. 
% Formally, if we take a vector operation of the dot-product between two vectors, the result of the operation is zero.
% In \autoref{eq:dot_product}, we sampled one query result from the first row of \autoref{tab:CQ1}. 
% The result of the dot product operation is zero, i.e., the configuration of the slip plane normal, $\vec{n}_0$, and the slip direction, $\vec{s}_0$, in slip\_sys\_0 is correct.
% \begin{equation}
% \label{eq:dot_product}
%      \vec{n}_0\cdot\vec{s}_0=
%   \begin{pmatrix} -1 & -1 & 1 \end{pmatrix} 
%   \begin{pmatrix} 0\\ -1 \\ -1 \end{pmatrix} = 0
% \end{equation}
% Furthermore, each slip system consists of a slip plane normal and a slip direction.
The complete set of the competency questions and the corresponding SPARQL queries can also be found in the DISO GitHub repository\footnote{\url{https://github.com/Materials-Data-Science-and-Informatics/Dislocation-Ontology-Suite/blob/main/DISO/CQs/CQs.md}}.

% \begin{itemize}
%     \item CQ4: What are the slip systems of a given crystal structure?
% \end{itemize}
% For the complete CQs, please visit the \href{https://github.com/Materials-Data-Science-and-Informatics/dislocation-ontology/blob/master/CQs/CQs.md}{DISO GitHub repository}

% \subsection{Corresponding SPARQL Query}

% The complete set of the competency questions and the corrsponding SPARQL queries can be found in the \href{https://github.com/Materials-Data-Science-and-Informatics/dislocation-ontology/blob/master/CQs/CQs.md}{DISO GitHub repository}.

\subsection{OntoQA Evaluation}
In this section, we assess the richness of DISO by using a criteria-based evaluation called OntoQA~\cite{tartir200}. OntoQA evaluates the ontology using schema metrics and population/instance metrics. To evaluate the ontology design and its potential for rich knowledge representation, we use the following metrics:
\begin{itemize}
    \item \emph{Relationship richness (\textbf{RR})} shows the diversity of relations and placement of relations in the ontology. Formally, it is defined by 
    \begin{equation}
        RR = \frac{|P|}{|SC|+|P|}
    \end{equation} where \textbf{P} is the number of relationships in the ontology and \textbf{SC} is the number of sub-classes. The more relations the ontology owns, except \emph{is-a} relations, the richer it is. 
    \item \emph{Attribute richness (\textbf{AR})} shows the more slots/attributed that are defined, the more knowledge the ontology delivers. Formally, AR  is defined by 
    \begin{equation}
        AR = \frac{|AT|}{|C|}
    \end{equation} where \textbf{AT} is the number of attributes for all classes and \textbf{C} is the number of classes.
    
    \item \emph{Inheritance richness (\textbf{IR})} describes the distribution of information across different levels of the ontology inheritance tree. IR indicates how knowledge is grouped into different classes and sub-classes in the ontology. Formally, IR is defined by 
    \begin{equation}
        IR = \frac{|SC|}{|C|}
    \end{equation} 
    % A high number of IR the vertical ontology, i.e., the ontology contains a large number of inheritance levels.
\end{itemize}

As shown in \autoref{tab:ontoqa}, we compare the results of DISO with MDO~\cite{li2020} and CSO\footnote{\url{https://purls.helmholtz-metadaten.de/disos/cso}}.
The former is an ontology representing the domain of solid-state physics simulation in materials science. 
The latter represents crystallographic information needed to describe the dislocation. 
While the MDO ontology has a larger \textit{RR}, i.e., it has more diversity in relations in the ontology. 
DISO and CSO have similar values in terms of \textit{IR}, which means they represent a wider range of knowledge compared to MDO. 
Concerning \textit{AR}, DISO has a larger \textit{AR} than MDO but is slightly smaller than CSO, enabling more knowledge per instance, which is more useful in the DD data.
% it enables more knowledge the ontology delivers, and it represents how well knowledge is grouped into different categories, respectively.

\begin{table}[H]
    \centering
    \caption{OntoQA Evaluation of DISO}
    \begin{tabular}{@{}cccccccc@{}} 
 \hline
 Ontology & C & SC & AT & P & RR & AR & IR \\ [0.5ex] 
 \hline\hline
 MDO & 37 & 49 & 14 & 32 & 0.40 & 0.38 & 1.33\\
 \hline
 CSO & 30 & 49 & 19 & 24 & 0.33 & 0.63 & 1.63\\
 \hline
 DISO & 38 & 62 & 23 & 33 & \textbf{0.35} & \textbf{0.61} & \textbf{1.63} \\
 \hline
\end{tabular}
\label{tab:ontoqa}
\end{table}

\section{Conclusion and Outlook}
\label{sec6/conclusion}
In this paper, we presented the dislocation ontology, which is an ontology that defines the linear defect concepts and relations in crystalline materials. 
% Moreover, we introduce CSO and CDO which are ontologies defining the concepts of crystal structure and crystallographic defect, respectively. 
% Moreover, we introduced two ontologies (i.e. CSO and CDO) that define crystal structure and crystallographic defect concepts.
% The ontology development processes are described following several practices, e.g., adding ontology metadata and reusing existing ontologies, e.g., MDO, EMMO, and QUDT. 
%Lastly, we evaluated DISO by answering competency questions through SPARQL queries explained.
DISO is developed using a top-down approach and is published through a persistent identifier, following W3C best practices for publishing Linked Data.
Two potential use cases for DISO are presented to illustrate its usefulness in the dislocation dynamics domain.
We followed two evaluation strategies to demonstrate the usefulness and the quality of the ontology.
We developed Python scripts to annotate the dislocation data generated from the DDD MoDELib software with DISO classes resulting in an RDF dataset.

We are planning to improve DISO by modeling the linear elasticity theory of dislocations, other defects (e.g., point defects, grain boundaries), and the application use cases in the dislocation domain, e.g., TEM data and dislocation simulation.
In addition, ontology alignment into EMMO would also be a very worthwhile undertaking. This alignment extends the interoperability between domain ontologies, particularly within the MSE community.
The ontology will continue to be maintained and extended in the context of the Helmholtz Metadata Collaboration efforts of facilitating machine readability and reuse of research data.

Since modularity is considered an emerging approach for developing ontologies, we are in the process of developing a dislocation ontology suite (DISOS) which comprises several modules each representing concepts in a specific area in the dislocation dynamics domain.

\section{Acknowledgments}
\begin{sloppypar}
AI and SS acknowledge financial support from the European Research Council through the ERC Grant Agreement No. 759419 MuDiLingo ("A Multiscale Dislocation Language for Data-Driven Materials Science").
AI, SF, and SS acknowledge Helmholtz Metadata Collaboration (HMC) within the Hub Information at the Forschungszentrum Jülich (FZJ).

Finally, we are grateful to Aytekin Demirci for the MoDELib parser that we used.
\end{sloppypar}
\bibliographystyle{ACM-Reference-Format}
\bibliography{references} 
\end{document}